\documentstyle[12pt,epsfig]{article}
\def\fun#1#2{\lower3.6pt\vbox{\baselineskip0pt\lineskip.9pt
  \ialign{$\mathsurround=0pt#1\hfil##\hfil$\crcr#2\crcr\sim\crcr}}}
\newcommand{\dd}{\mbox{d}}
\newcommand{\vecc}[1]{\mbox{\boldmath $#1$}}

\title{\bf \huge Small--Angle Electron-Positron Scattering \bf \Large $^\dagger$} 

\author{A.B.~Arbuzov$^{a}$, V.S.~Fadin$^{b}$,  
E.A.~Kuraev$^{a}$, L.N.~Lipatov$^{c}$, \\
N.P.~Merenkov$^{d}$ and L.G.~Trentadue$^{e}$}

\date{}

\begin{document}

\maketitle

\vskip 0.3cm
\begin{center}
{$^a$ \it Joint Institute for Nuclear Research,
Dubna, Moscow region, 141980, Russia \\
$^b$ \it Budker Institute for Nuclear Physics and Novosibirsk State University,
630090, Novosibirsk, Russia \\
$^c$ \it St.-Petersburg Institute of Nuclear Physics, 
Gatchina, Leningrad region, 188350, Russia \\
$^d$ \it Physico-Technical Institute,
Kharkov,  310108,  Ukraine\\
$^e$ \it Dipartimento di Fisica, Universit\'a di Parma and 
INFN, Gruppo Collegato di Parma,\\ 43100 Parma, Italy }
\end{center}

\vskip 3.0cm
\begin{abstract}
We consider small--angle electron--positron scattering in 
Quantum Electrodynamics.  
Leading logarithmic contributions to the cross--section are explicitly 
calculated 
to three loop. Next--to--leading terms are exactly computed to
two loop. 
All the radiative corrections due to photons
as well as pair production are taken into account.
The impact of newly evaluated next-to-leading and higher order leading
corrections is discussed and numerical results are explicitly given.
The results obtained are generally valid for high
and low energy $e^+e^-$ colliders.
At LEP and SLC these results can be used to
reduce the uncertainty on the cross--section 
below the per mille level. \\[.4cm]
PACS numbers  ~~12.15.Lk, ~12.20.--m, ~12.20.Ds, ~13.40.--f
\end{abstract}

\vfill
\begin{flushleft}
University of Parma \\
preprint UPRF-96-474
\end{flushleft}
\vspace{1cm}
$^\dagger${\footnotesize Work supported by the Istituto Nazionale di Fisica Nucleare (INFN).
INTAS Grant 1867-93.}
\newpage
An high accuracy, better than one {\em per mille,\/} 
measurement of the luminosity has been reached at LEP
\cite{rs1}. 
The small--angle electron--positron scattering (Bhabha) process 
is normally used as the reference cross--section to measure the
luminosity.

A poorly known Bhabha cross--section has the consequence 
of producing a systematic error on the 
determination of relevant physical observables as, for example,
the hadronic peak cross--section $\sigma^h_{peak}$ or the leptonic widths 
$\Gamma_{e,\mu}$. This poor accuracy reflects itself on the extraction
of the Standard Model parameters 
on the exclusion of new physics signals as, for example, those coming
from a deviation from the Standard Model prediction
for the number of light neutrinos \cite{rs1}.
The experimental uncertainty in the luminosity determination together 
with the theoretical one define the
uncertainties on the above mentioned quantities.
An adequate
theoretical accuracy 
on the cross--section is therefore highly needed. 

Various approaches have been used to obtain the Bhabha 
cross--section \cite{rs2,rs3}.
Some of them (see ref.~\cite{rs2}) are based on Monte Carlo 
generators. Others use the structure function method to calculate
 the radiative 
corrections \cite{rs3}. With both methods the contributions
of next--to--leading  as well as higher order
radiative corrections have not been 
systematically evaluated
and this yielding an important source of uncertainty above the level
of $\delta \sigma / \sigma \simeq 0.001$ accuracy.
Therefore an equally accurate theoretical 
determination of the Bhabha cross--section, even if it is approaching to,
has not yet reached the experimental accuracy \cite{rs1a}.

In this letter we describe the results obtained with a
different approach based on the direct evaluation of Feynman 
diagrams. The results, given in analytical form, systematically take into 
account leading as well as next--to--leading contributions 
thus reducing the physical uncertainty to the $0.1\%$ level.

The differential cross--section for the small-angle Bhabha process
in the Standard Model is \cite{rs4}:
\begin{eqnarray} \label{born}
\frac{\dd\sigma_B}{\theta \dd\theta}
&=&\frac{8\pi\alpha^2}{\theta^4}(1+\delta_{\theta}+\delta_{\mbox{weak}}), \quad
\theta \ll 1,\quad
\delta_{\theta}=-\theta^2/2 + 9\theta^4/40, \\ \nonumber
\delta_{\mbox{weak}}&=&2g_v^2\xi - \frac{\theta^2}{4}(g_v^2+g_a^2)
\Re\mbox{e}\;\chi + \frac{\theta^4}{32}(g_v^4+g_a^4+6g_v^2g_a^2)
|\chi|^2, \\ \nonumber
\chi&=&s[(s-M_Z^2+iM_Z\Gamma_Z)\sin(2\theta_W)]^{-1},\quad
\xi=t[(t-M_Z^2)\sin(2\theta_W)]^{-1},
\end{eqnarray}
where $\theta$ is the scattering angle, $s=4E^2$, $t=-E^2\theta^2$,
~$E$ is the beam (center-of-mass) energy, $\theta_W$ the Weinberg angle 
and $g_a=-1/2$, $g_v=-(1-4\sin^2\theta_W)/2$ the axial and vector couplings
of the $Z^0$ boson.
Experimental cross-sections are obtained by collecting
events within particular angular portions of the detectors with given
energy cuts. To compare with those observed distributions one has to take 
into account radiative corrections to higher orders in perturbation theory 
and it is convenient to introduce the dimensionless quantity:

\begin{eqnarray} \label{Sigma}
\Sigma=\frac{Q_1^2}{4\pi\alpha^2}\int\limits_{0}^{1}\dd x_1
\int\limits_{0}^{1}\dd x_2 \Theta(x_1x_2-x_c)
\int\dd^2\vecc{q}^{\bot}_1\Theta^c_1
\int\dd^2\vecc{q}^{\bot}_2 \Theta^c_2 
\frac{\dd 
\sigma^{e^+e^-\rightarrow e^+(\vecc{q}^{\bot}_2,x_2)
e^-(\vecc{q}^{\bot}_1,x_1)+X}}{\dd x_1 d^2 \vecc{q}^{\bot}_1 
\dd x_2 \dd^2 \vecc{q}^{\bot}_2}\, ,
\end{eqnarray}
where $x_c$ is the energy fraction threshold for the detection
of the final electron and positron $x_c\leq x_1x_2$, where $x_{1,2}$ are the
energy fractions of the scattered leptons and $\vecc{q}^{\bot}_1$,
$\vecc{q}^{\bot}_2$ are the components of their momenta transverse with
respect to the beam directions. $\Theta^c_{1,2}$
are the step functions which account for the registration of the scattered 
leptons by the circular detectors situated close to the beams with an 
aperture bounded by the angles $\theta_{1,2,3,4}$:
\begin{eqnarray} \label{Tet}
\Theta^c_1=\Theta(\theta_3-\frac{|\vecc{q}^{\bot}_1|}{x_1E})
\Theta(\frac{|\vecc{q}^{\bot}_1|}{x_1E}-\theta_1), 
\qquad 
\Theta^c_2=\Theta(\theta_4-\frac{|\vecc{q}^{\bot}_2|}{x_2E})
\Theta(\frac{|\vecc{q}^{\bot}_2|}{x_2E}-\theta_2).
\end{eqnarray}
If $\theta_1$ is the minimal aperture angle we define the minimal
momentum transferred from the electron to the positron as $Q_1^2=E^2\theta_1^2$.
At LEP/SLC $Q_1 \simeq 1~GeV/c$. For symmetric detectors:
$\theta_1=\theta_2$, $\theta_3=\theta_4$, $\rho=\theta_3/\theta_1$.
\begin{eqnarray} \label{SS}
\Sigma&=&\Sigma_0+\Sigma^{\gamma}+\Sigma^{\gamma\gamma}
+\Sigma^{\gamma}_{\gamma}+\Sigma^{e^+e^-}+\Sigma^{3\gamma}
+\Sigma^{e^+e^-\gamma} \\ \nonumber
&=&\Sigma^0_0(1+\delta_{\mbox{tot}})=\Sigma^0_0(1+\delta_0+\delta^{\gamma}
+ \delta^{2\gamma} + \delta^{e^+e^-} + \delta^{3\gamma} + \delta^{e^+e^-\gamma}) \\ \nonumber
&=& \Sigma^0_0(1+\sum\limits_{i}\delta^i), \qquad
\delta^{2\gamma}\equiv \delta^{\gamma\gamma}+\delta^{\gamma}_{\gamma},
\qquad \Sigma^0_0=1-\frac{1}{\rho^2}\, ,
\end{eqnarray}
$\Sigma_0$  has the form:
\begin{eqnarray} \label{S0}
\Sigma_0=\int\limits^{\rho^2}_{1}\frac{\dd z}{z^2}
U(-zQ_1^2)(1+\delta_{\mbox{weak}}(\theta^2)
+\delta_{\theta}(\theta^2))\bigg|_{\theta^2=z\theta_1^2}
\end{eqnarray}
with $U(t)=(1-\Pi(t))^{-2}$ where $\Pi(t)$ is the vacuum 
polarization operator of the exchanged photon \cite{rs5}.
The experimentally observable cross--section is given by the expression:
\begin{eqnarray} \label{sigma}
\sigma=\frac{4\pi \alpha^2}{Q_1^2}\Sigma^0_0(1+\delta_0+\delta^{\gamma}
+ \delta^{2\gamma} + \delta^{e^+e^-} + \delta^{3\gamma} + \delta^{e^+e^-\gamma}).
\end{eqnarray}

The quantity $\Sigma^{\gamma}$ collects the radiative corrections to the Born
amplitude related to the emission of single virtual, real soft or hard photon.
Virtual and soft photon contributions to the differential cross--section 
are proportional to the Born amplitude:
$\dd\sigma=\dd\sigma_B(1+\frac{\alpha}{\pi}(2(L-1)
\ln\Delta+3L/2-2))$. This quantity contains the {\em large logarithm\/} $L$,
$L=\ln(zQ_1^2/m^2)$ ($m$ is the electron mass), and 
$\Delta=\delta E/E$, where $\delta E$ is the maximal energy 
carried by a soft photon, $\Delta\ll 1$. The $\Delta$--dependence
disappears in the total sum when the emission 
of a hard photon with energy fraction larger than $\Delta$ 
is also taken into account.

A simple analytical expression for the small--angle Bhabha
differential cross--section for 
one hard photon emission  can be written within the 
the infinite-momentum  frame formalism. One obtains:
\begin{eqnarray} \label{sh}
&& \frac{\dd\sigma^H}{\dd x\dd^2 \vecc{q}^{\bot}_1 
\dd^2\vecc{q}^{\bot}_2}=\frac{2\alpha^2}{\pi^2}
\frac{1+x^2}{((\vecc{q}^{\bot}_2)^2)^2(1-x)}
\biggl[\frac{(\vecc{q}^{\bot}_2)^2(1-x)^2}{d_1d_2}-
\frac{2m^2x(1-x)^2(d_1-d_2)^2}{(1+x^2)d_1^2d_2^2}\biggr], \\ \nonumber
&& d_1=m^2(1-x)^2+(\vecc{q}_1^{\bot} - \vecc{q}^{\bot}_2)^2=(1-x)2p_1k, 
\\ \nonumber
&& d_2=m^2(1-x)^2+(\vecc{q}_1^{\bot} - x\vecc{q}^{\bot}_2)^2=x(1-x)2q_1k, \quad
\vecc{q}^{\bot}_2=\vecc{q}^{\bot}_1+\vecc{k}^{\bot}, \\ \nonumber
&& x=\frac{q_1^0}{p_1^0}\, , \quad
q_1=xp_1+\vecc{q}_1^{\bot}, \quad
p_1+p_2=q_1+k+q_2\, ,
\end{eqnarray}
where $p_1$, $p_2$ and $k$ are the 4-momenta of the initial
electron, positron and emitted photon respectively,
$x$ and $1-x$ are the energy fractions of the scattered electron,
$\vecc{k}^{\bot}$ is the transverse component of the photon momentum 
with respect to the beam direction.
The photon is supposed to be emitted from the electron line.
The same result is obtained for the emission from the positron line.
The above expression takes
into account only scattering--type diagrams. 
According to the definition of eq.(\ref{Sigma}) we have that
\begin{eqnarray} \label{S1g}
\Sigma^{\gamma}&=&\frac{\alpha}{\pi}\int\limits_{1}^{\rho^2}
\frac{\dd z}{z^2}\int\limits_{x_c}^{1}
\dd x U(-zQ_1^2)\;\biggl\{
(L-1) P(x)[1+\Theta(x^2\rho^2-z)] \\ \nonumber
&+& \frac {1+x^2}{1-x}k(x,z)- \delta(1-x) \biggr\}, \\ \nonumber
k(x,z)&=& \frac{(1-x)^2}{1+x^2} \;[1+\Theta(x^2\rho^2-z)]
+ \; L_1+\Theta(x^2\rho^2-z)\;L_2
+\; \Theta(z-x^2\rho^2) L_3 \, ,
\end{eqnarray}
\begin{eqnarray*} 
L_1=\ln\left|\frac{x^2(z-1)(\rho^2-z)}{(x-z)(x\rho^2-z)}\right|, \ \
L_2=\ln\left|\frac{(z-x^2)(x^2\rho^2-z)}{x^2(x-z)(x\rho^2-z)}\right|, \ \
L_3=\ln\left|\frac{(z-x^2)(x\rho^2-z)}{(x-z)(x^2\rho^2-z)}\right|,
\end{eqnarray*}
and
\begin{eqnarray} \label{p1}
P(x)=\biggl(\frac{1+x^2}{1-x}\biggr)_+=\lim_{ \Delta \to 0 }\;
\biggl\{ \frac{1+x^2}{1-x}\;
\Theta(1-x-\Delta)+(\frac{3}{2}+2\ln\Delta)\;\delta(1-x) \biggr\}
\end{eqnarray}
is the kernel of the non--singlet (NS) evolution equation \cite{ee}.

For small scattering angles $(-t/s\sim \theta^2\ll 1)$
the elastic amplitude $A(s,t)$
has a generalized eikonal form:
$A(s,t)=A_{\mbox{Born}}(t)\;\Gamma^2(t)\;\sqrt{U(t)}\;\mbox{exp}\{i\phi(t)\}$, 
where $\Gamma(t)$ is the Dirac electron form
factor and $\phi(t)$ is the Coulomb phase \cite{rs6}. The Coulomb phase factor collects
the multiple exchange photon contributions in the scattering channel.
This holds for elastic as well as for inelastic amplitudes.
This form permits us to neglect diagrams with several exchanged photons 
since the phase factor $\mbox{exp}\{i\phi(t)\}$ 
disappears in the physical cross-sections.
To higher orders in the perturbative expansion, only Feynman diagrams with 
one exchanged photon between electron and positron line do contribute.
For the second order photonic contributions we use the
known results of the electron Dirac form factor\cite{rs7}. Pauli form factor
gives a negligible contributions of ${\cal O}(\theta^2(\alpha/\pi)^2)$. 
They include the cross-section for the emission of two soft photons
as well as the emission a single real (soft or hard) photon
with one--loop virtual radiative corrections \cite{a} 
and the cross--section for the emission of two hard photons in the same
or in opposite directions. For this last kinematical configuration
we may use a generalization of the single bremsstrahlung
cross-section in eq.(\ref{sh}) \cite{b}. We obtain the following expression:
\begin{eqnarray} \label{Sgug}
\Sigma_{\gamma}^{\gamma}&=&\frac{1}{4}\bigl(\frac{\alpha}{\pi}\bigr)^2
\int\frac{\dd z}{z^2}
U(-Q_1^2z)\Biggl\{ L^2\int\limits_{x_c}^{1}\dd x_1
\int\limits_{x_c/x_1}^{1}\dd x_2 P(x_1)P(x_2) \\ \nonumber
&\times& \bigl[\Theta(z-1)\Theta(\rho^2-z) 
+ \Theta(z-x_1^2)\Theta(x_1^2\rho^2-z)\bigr] \\ \nonumber
&\times& \bigl[\Theta(z-1)\Theta(\rho^2-z)
+ \Theta(z-x_2^2)\Theta(x_2^2\rho^2-z)\bigr] \;
+\;  L\Phi^{\gamma}_{\gamma} \Biggr\},
\end{eqnarray}
with $\Phi^{\gamma}_{\gamma}$ an analytical function (see \cite{rs10}).
To calculate within the single logarithmic accuracy the contribution 
of the emission of two hard photons from a single lepton line (the electon 
in our case) we separate, by introducing a new auxiliary
parameter $\theta_0$  $(1\gg \theta_0\gg m/\varepsilon)$,
in collinear and 
semi-collinear kinematical regions \cite{c}. Namely, in 
the collinear region both emitted photons move in the narrow cones
defined by the polar angle $\theta_0$ with respect to the initial or final
(as well as simultaneously) electron 3-momenta. This region gives 
leading and next--to--leading contributions. The semi--collinear
region produces only next--to--leading terms and corresponds to the
kinematical configurations when only one of the photons moves inside one of the
defined narrow cones when the second is radiated outside.
It can be shown that, as it should, the dependence on $\theta_0$ 
disappears in the total sum:
\begin{eqnarray} \label{Sgg}
\Sigma^{\gamma\gamma}&=&\frac{1}{2}\bigl(\frac{\alpha}{\pi}\bigr)^2
\int\limits_{1}^{\rho^2}\frac{\dd z}{z^2}
U(-Q_1^2z)\Biggl\{L^2\int\limits_{x_c}^{1}\dd x\;\Bigl\{
\frac{1}{2}P^{(2)}(x)\;[\;\Theta(x^2\rho^2-z)+1] \\ \nonumber
&+& \int\limits_{x}^{1}\frac{\dd t}{t}
P(t)\;P(\frac{x}{t})\;\Theta(t^2\rho^2-z)\Bigr\}
\; +\; L\Phi^{\gamma\gamma} \Biggr\},  \\  \label{p2}
P^{(2)}(x)&=&\int\limits_{x}^{1}\frac{\dd t}{t}P(t)\;P(\frac{x}{t})=
\lim_{\Delta \to 0} \biggl\{\;\biggl[\biggl(2\ln\Delta+\frac{3}{2}\biggr)^2
- 4\zeta_2\biggr]\;\delta(1-x) \\ \nonumber 
&+& 2\biggl[\frac{1+x^2}{1-x}\biggl(2\ln(1-x)-\ln x + \frac{3}{2}\biggr)
+\frac{1}{2}(1+x)\ln x- 1+x\biggr]\;\Theta(1-x-\Delta)\biggr\}.
\end{eqnarray}
For $\Phi^{\gamma\gamma}$ see \cite{rs10}.
To the same order of perturbation theory we have to take into account the 
$e^+e^-$ pair production processes.
For hard pair production we
again consider four collinear kinematical regions,
when the created pair moves close to the directions of the projectiles
or of the scattered particles, and six semi--collinear regions which
having $2\to 3$ like kinematics
i.e. when one of the final particles moves 
close to the beam direction or when one component of the created
pair is close the other.
The cancellation of the auxiliary parameter $\theta_0$
can be explicitly verified \cite{rs8}.
 
The relevant two--loop contribution from the
electron form factor \cite{rs7} contains $L^3$ terms which disappear
in the sum with the contribution due to soft pair production \cite{rs8}:
\begin{eqnarray} \label{See}
\Sigma^{e^+e^-}&=&\frac{1}{4}\bigl(\frac{\alpha}{\pi}\bigr)^2
\int\limits_{1}^{\rho^2}
\frac{\dd z}{z^2} U(-zQ_1^2)\biggl\{ L^2
\int\limits_{x_c}^{1}\dd x [1+\Theta(z-1)\Theta(x^2\rho^2-z)]\;R(x)
\; +\; L\Phi^{e^+e^-}\biggr\}, \\ \nonumber
R(x)&=&\frac{2}{3}P(x) + 2(1+x)\ln\;x + \frac{1-x}{3x}
(4+7x+4x^2),
\end{eqnarray}
where $\Phi^{e^+e^-}$ collects nonleading corrections which can be found in
\cite{rs10}.

Within the third order of perturbation theory it is sufficient,
to the  accuracy of the ${\cal O}(10^{-3})$, to
consider only leading contributions related to the initial particle
radiation. Leading contributions can be systematically included by using QED
evolution equations \cite{ee}. 

We consider the channels of $\gamma\gamma\gamma$ and 
$\gamma e^+e^-$ production real or virtual in all possible
combinations. We obtain
\begin{eqnarray} \label{S3g}
\Sigma^{3\gamma}&=&\frac{1}{4}\;(\frac{\alpha}{\pi}{\cal{L}})^3
\int\frac{\dd z}{z^2} \int\limits_{x_c}^{1}\dd x_1 \int\limits_{x_c/x_1}^{1}
\dd x_2\; \biggl[\frac{1}{6}\delta(1-x_2)\;P^{(3)}(x_1)
\\ \nonumber 
& \times & \Theta(x_1^2\rho^2-z)\Theta(z-1)
+\frac{1}{2}P^{(2)}(x_1)P(x_2)\Theta_1\Theta_2\biggr], \\ \nonumber
\Theta_{1,2}& \equiv & \Theta(z-x_{1,2}^2)\Theta(\rho^2x_{1,2}^2-z),
\\ \label{p3}
P^{(3)}(x)&=&\int\limits_{x}^{1}\frac{\dd y}{y}P^{(2)}(y)P(\frac{x}{y}),
\qquad {\cal L}=\ln\frac{Q_1^2}{m^2}\, ,
\\ \label{Seeg}
\Sigma^{e^+e^-\gamma}&=&\frac{1}{4}(\frac{\alpha}{\pi}
{\cal{L}})^3\int\frac{\dd z}{z^2}
\int\limits_{x_c/x_1}^{1}\dd x_1\int\limits_{x_c}^{1}\dd x_2
\biggl\{\frac{1}{3}[R^P(x_1)-\frac{1}{3}R^s(x_1)]
\\ \nonumber &\times&
\delta(1-x_2)\Theta(x_1^2\rho^2-z)\Theta(z-1)+\frac{1}{2}\;P(x_2)R^r(x_1)
\;\Theta_1\Theta_2\biggr\},
\end{eqnarray}
where
\begin{eqnarray}
R^r(x)&=&R^s(x)+\frac{2}{3}P(x), \qquad
R^s(x)=\frac{1-x}{3x}(4+7x+4x^2)+2(1+x)\ln x,
\\ \nonumber 
R^P(x)&=&R^s(x)\biggl(\frac{3}{2}+2\ln(1-x)\biggr)
+ (1+x)\biggl( - \ln^2x - 4\int\limits_{0}^{1-x}\dd y
\frac{\ln(1-y)}{y}\biggr)
\\ \nonumber 
&+&\frac{1}{3}(-9-3x+8x^2)\ln x+
\frac{2}{3}\biggl(-\frac{3}{x}-8+8x+3x^2\biggr)+\frac{2}{3}P^{(2)}(x).
\end{eqnarray}

By combining the partial results in eqs.(\ref{S0}),(\ref{S1g}),(10--14),
and (\ref{Seeg}) one obtains the final result 
for the observable cross--section eq.(\ref{sigma}).

In Table~1 we give our results for different values
of the threshold energy fraction $x_c$ for the defined angular acceptance
of $\theta_1=1.6^{\circ}$ and $\theta_2=2.8^{\circ}$. 
Nonleading second order photonic corrections are negative and
turn out to be larger in magnitude than both third order 
and second order ones due to pair production. This is shown in Fig.~1. 
Photonic corrections
(leading and nonleading) (see Fig.~2) due to double photon emission
from both fermions dominate the ones from a single lepton line. 
This fact is a consequence
of the sign--changing $P^{(2)}(x)$ function
(see eq.(\ref{p2})), which describes
double bremsstrahlung from a single lepton line.
The importance of radiative corrections to Bhabha cross--section
is illustrated in Fig.~3. The decreasing of the cross--section in the region 
$x_c \to 1$ can be understood as a reduction by increasing 
$x_c$ of the positive contribution
of real photon emission, while first order virtual corrections,
being negative, remain unchanged.

The accuracy of this result is implicitely defined by the 
terms omitted in the perturbative series. Typically they are of the type:
\begin{eqnarray}
\frac{\alpha}{\pi}\theta^2,\ \bigl(\frac{\alpha}{\pi}\bigr)^2,\
\bigl(\frac{\alpha}{\pi}\bigr)^3L^2,\ \bigl(\frac{\alpha}{\pi}L\bigr)^4. 
\end{eqnarray}
An estimate of their magnitudes permits us to state that the uncertainty
of our result to be at the level $10^{-4}$. A detailed analysis of the
attained accuracy is given in \cite{rs10}.
From the analysis given above clearly emerges the importance of the 
next--to--leading
contributions to reach an accuracy adequate to the one attained
in present LEP and SLC experiments. The above formulae can be applied to 
future high--energy $e^+e^-$ colliders as well.

\vskip 10pt
We are grateful for support to the Istituto Nazionale 
di Fisica Nucleare (INFN), to the International Association 
for the Promotion of Cooperation with Scientists (INTAS) 
for the grant 93-1867 and 
to the Russian Foundation for Fundamental Investigations (RFFI)
for the grant 96-02-17512.
One of us (L.T.) would like to thank H.~Czyz, M.~Dallavalle,
B.~Pietrzyk and T.~Pullia for several useful discussions at various stages
of the work and the CERN theory group for the hospitality. 
Three of us (A.A., E.K. and N.M.) would like to thank the INFN
Laboratori Nazionali
di Frascati, the Dipartimento di Fisica dell'Universit\'a di Roma
"Tor Vergata" and the Dipartimento di Fisica
dell'Universit\'a di Parma for their hospitality
during the preparation of this work.
One of us (A.A.) is thankful to the Royal Swedish Academy of Sciences for
an ICFPM grant.

\newpage
\section{Figure Captions}

\vskip 1.0pt \noindent
Fig.1: The ratios a) $\delta^{2\gamma}_{nonleading}
/\delta_{tot}$, b) $\delta^{e^+e^-}/\delta_{tot}$ and c)
$(\delta^{3\gamma}+\delta^{e^+e^-\gamma})/\delta_{tot}$
as functions of $x_c$ for $\theta_1=1.6^{\circ}$, 
$\theta_2=2.8^{\circ}$ and $\sqrt{s}=91.161$.

\vskip 3.0pt \noindent
Fig.2: Behaviour of a) $\delta^{\gamma\gamma}_{leading}$ (solid line)  
and $\delta^{\gamma\gamma}_{leading}+\delta^{\gamma\gamma}_{nonleading}$
(dashed line), b) ${\delta^{\gamma}_{\gamma}}_{leading}$ (solid line)
and ${\delta^{\gamma}_{\gamma}}_{leading}
+{\delta^{\gamma}_{\gamma}}_{nonleading}$
(dashed line) as functions of $x_c$ (parameters are as in Fig.1).

\vskip 3.0pt \noindent
Fig.3:  Corrected cross-section of Bhabha scattering according 
to eq.(\ref{sigma})
as a function of $x_c$: a) $\sigma_{0}$+${\cal O}(\alpha)$ corrections,
b) with ${\cal O}(\alpha^2L^2)$ photonic corrections added,
c) with all other corrections (see eq.(\ref{SS})) added.
$\sigma_{0}$ is the Born cross--section. 
For the parameters of as in Fig.1 one has $\sigma_{0}=106.33$~nb. 

\vspace{1.5cm}

\vspace{.3cm}
Table~1:  Per cent values of $\delta^i$ as defined in eq.(\ref{SS}) 
for $\sqrt{s}=91.161$ GeV, $\theta_1=1.61^{\circ}$,
$\theta_2=2.8^{\circ}$, $\sin^2\theta_W=0.2283$,
$\Gamma_Z=2.4857$ GeV. Quantity $\delta^{2\gamma}$
is equal to the sum $\delta^{\gamma\gamma}+\delta^{\gamma}_{\gamma}$
(see eq.(\ref{SS})).
\vspace{.3cm}

\begin{tabular}{|c|c|c|c|c|c|c|c|c|}
\hline
$x_c$ & $\delta_0 $ & $\delta^{\gamma} $
&$\delta^{2\gamma}_{\mbox{\tiny leading}}$
&$\delta^{2\gamma}_{\mbox{\tiny nonleading}}$
& $\delta^{e^+e^-} $ & $\delta^{e^+e^-\gamma} $
&$\delta^{3\gamma} $&$\sum \delta^i $ \\ \hline
0.1& 4.120& $-$8.918& 0.657&  0.162& $-$0.016& $-$0.017
& $-$0.019& $-$4.031$\pm$0.006 \\
0.2& 4.120& $-$9.226& 0.636&  0.156& $-$0.027& $-$0.011
& $-$0.016& $-$4.368$\pm$0.006 \\
0.3& 4.120& $-$9.626& 0.615&  0.148& $-$0.033& $-$0.008
& $-$0.013& $-$4.797$\pm$0.006 \\
0.4& 4.120&$-$10.147& 0.586&  0.139& $-$0.039& $-$0.005
& $-$0.010& $-$5.356$\pm$0.006 \\
0.5& 4.120&$-$10.850& 0.539&  0.129& $-$0.044& $-$0.003
& $-$0.006& $-$6.115$\pm$0.006 \\
0.6& 4.120&$-$11.866& 0.437&  0.132& $-$0.049& $-$0.002
& $-$0.001& $-$7.229$\pm$0.006 \\
0.7& 4.120&$-$13.770& 0.379&  0.130& $-$0.057& $-$0.001
&    0.005& $-$9.194$\pm$0.006 \\
0.8& 4.120&$-$17.423& 0.608&  0.089& $-$0.069&    0.001
&    0.013&$-$12.661$\pm$0.006 \\
0.9& 4.120&$-$25.269&1.952&$-$0.085& $-$0.085&    0.005
&    0.017&$-$19.379$\pm$0.006 \\
\hline
\end{tabular}

\vskip 20.0pt

\begin{figure}[t]
\begin{center}
\mbox{\epsfig{file=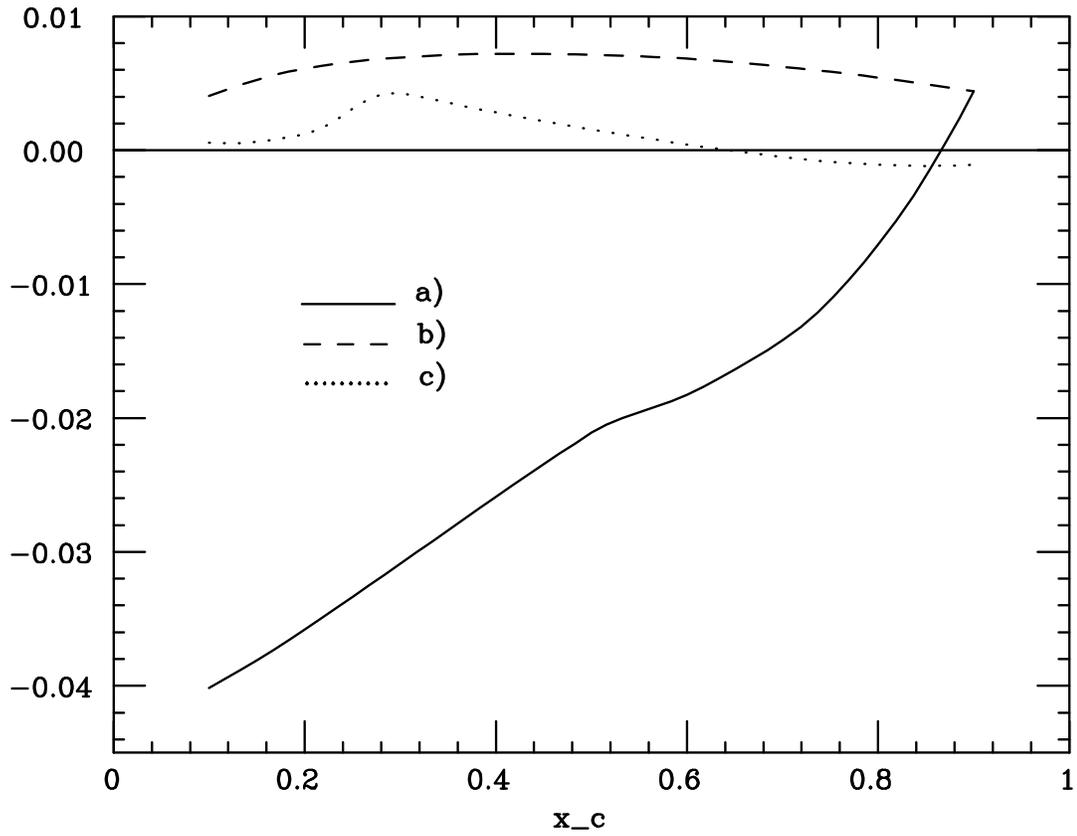,height=14cm,angle=90}}
\end{center}
\caption{
The ratios a) {$\delta^{2\gamma}_{nonleading}/\delta_{tot}$}, 
b){ $\delta^{e^+e^-}/\delta_{tot}$} and 
c) {$(\delta^{3\gamma}+\delta^{e^+e^-\gamma})/\delta_{tot}$} as functions of 
{$x_c$} 
for 
$s^{1/2}=91.161$~GeV 
and {$\theta_1=1.6^\circ$}, 
{$\theta_2=2.8^\circ$}.
}
\label{Fig1}
\end{figure}

\newpage

\begin{figure}[t]
\begin{center}
\mbox{\epsfig{file=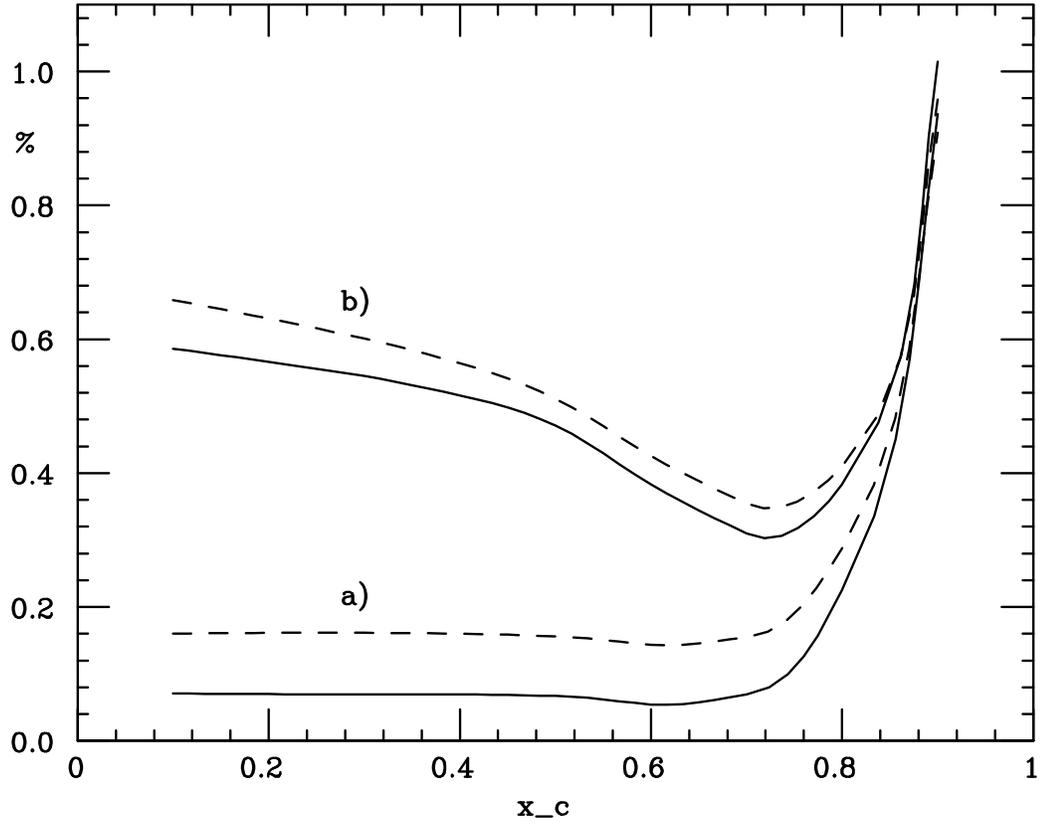,height=14cm,angle=90}}
\end{center}
\caption{
Behaviour of a) $\delta^{\gamma\gamma}_{leading}$ (solid line)  
and $\delta^{\gamma\gamma}_{leading}+\delta^{\gamma\gamma}_{nonleading}$
(dashed line), b) ${\delta^{\gamma}_{\gamma}}_{leading}$ (solid line)
and ${\delta^{\gamma}_{\gamma}}_{leading}
+{\delta^{\gamma}_{\gamma}}_{nonleading}$
(dashed line) as functions of $x_c$ (parameters are as in Fig.1).
}
\label{Fig2}
\end{figure}


\begin{figure}[t]
\begin{center}
\mbox{\epsfig{file=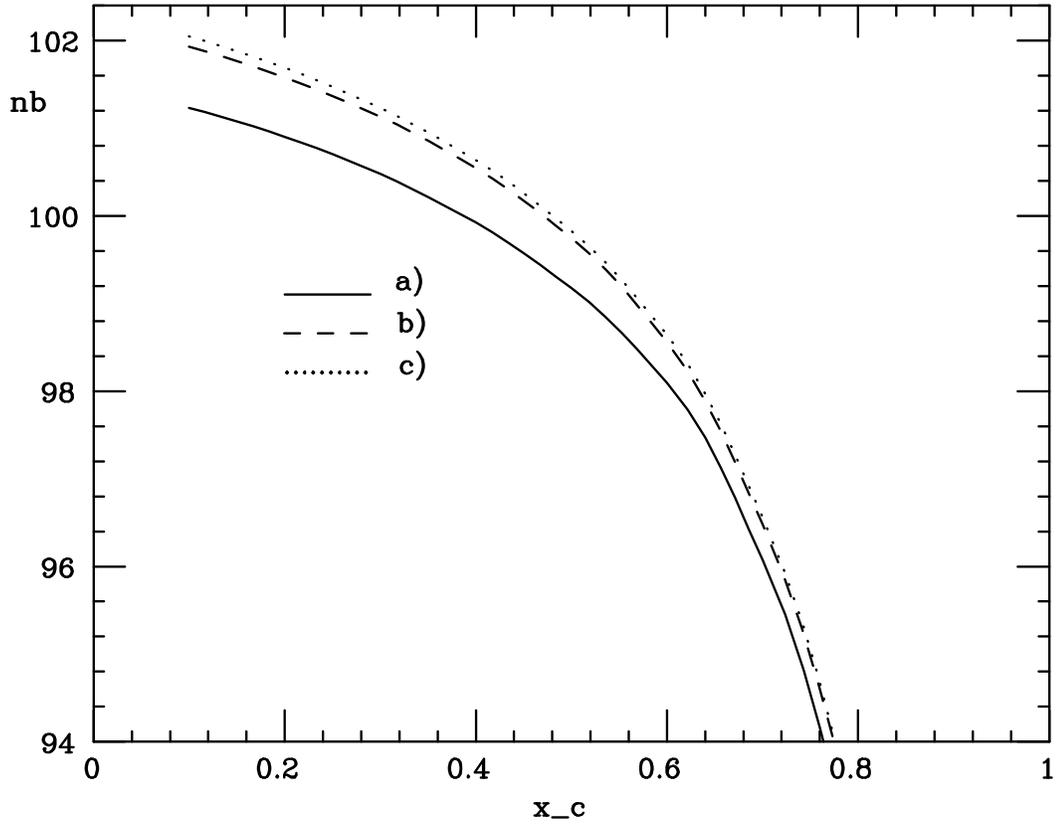,height=14cm,angle=90}}
\end{center}
\caption{Corrected cross--section of Bhabha scattering according 
to eq.(6)
as a function of $x_c$: a) $\sigma_0+{\cal O}(\alpha)$ corrections,
b) with ${\cal O}(\alpha^2L^2)$ photonic corrections added,
c) with all other corrections 
added, where
$\sigma_0$ is the Born cross--section. 
For the parameters of as in Fig.1 one has {$\sigma_{0}=106.33$}~nb.
}
\label{Fig3}
\end{figure}

\end{document}